\documentstyle[12pt]{article}
\begin{document}
%%%%%%%%%%%%%%%%%%%%%%%%%%%%%%%%%%%%%%%%%%%%%%%%%%%
\def\thefootnote{\fnsymbol{footnote}}
\begin{flushright}
KANAZAWA-96-03  \\ 
March, 1996
\end{flushright}
%\vspace{ .7cm}
\vspace*{2cm}
\begin{center}
{\LARGE\bf Flavor Structure of Soft SUSY Breaking Parameters}\\
\vspace{1 cm}
{\Large  Daijiro Suematsu}
\footnote[1]{e-mail:suematsu@hep.s.kanazawa-u.ac.jp}
\vspace {1cm}\\
{\it Department of Physics, Kanazawa University,\\
        Kanazawa 920-11, Japan}\\

{\it and}\\

{\it Theory Division, CERN, CH-1211 Geneve 23, Switzerland}    
\end{center}
\vspace{2cm}
{\Large\bf Abstract}\\  
%%%%%%%%%%%%%%%%%Abstract%%%%%%%%%%%%%%%%%%%%%%%%%%%
Flavor structure of soft supersymmetry breaking parameters is studied in a 
certain type of effective supergravity theory derived from moduli/dilaton 
dominated supersymmetry breaking.
Some interesting sum rules for soft scalar masses are presented.
They constrain their flavor structure and predict some interesting patterns 
appeared in
soft scalar masses in the non-universal case. We also study the alignment
phenomena in the flavor space of soft breaking parameters due to 
Yukawa couplings and discuss
their phenomenological consequences. 
\newpage
\setcounter{footnote}{0}
\def\thefootnote{\arabic{footnote}}
%%%%%%%%%%%%%%%%%Text%%%%%%%%%%%%%%%%%%%%%%%%%%%%%%%
\section{Introduction}
In supersymmetric theories soft supersymmetry breaking parameters play 
the crucial roles\cite{n}. Their phenomenological features are completely 
dependent on those parameters. This means that low energy phenomenology can
put the strong constraints on these parameters and may also
give some informations on the fundamental theory at high energy region,
which determines the structure of soft breaking parameters.
Especially, the constraints coming from the rare phenomena such as the 
flavor changing neutral current(FCNC)\cite{fcnc} and the electric dipole moment
of neutron(EDMN)\cite{edmn} are very strong. 
Because of these reasons the universality and reality of the soft
supersymmetry breaking 
parameters are usually assumed when the analysis of various phenomenological
aspects of the minimal supersymmetric standard model(MSSM) is done.
Up to now in many works the flavor structure of soft breaking
parameters has been discussed on the basis of suitable fundamental
frameworks\cite{dks}.
They mainly treat how the universality of soft scalar masses is realized
at low energy region. 

Recently it has been noticed that in superstring theory soft 
supersymmetry breaking parameters are generally non-universal\cite{il,kl,bim} 
and their various phenomenological consequences have been studied in that 
framework\cite{mn,ksy}.
These works show that non-universal soft breaking parameters bring rather 
different phenomenological features in comparison with the universal ones. 
In this situation
it seems to be very interesting to study the more detailed flavor structure
of these parameters on the basis of certain fundamental frameworks.
Flavor structure of soft breaking parameters seems not to have been studied 
enough beyond whether they are universal or not. This is, in part, because of
the above mentioned phenomenological constraints and also the predictivity 
of the theory.
However, there can be various flavor structures even if the phenomenological 
constraints are imposed. FCNC constraints, indeed, only require the
mass degeneracy among squarks with the same quantum numbers.
This point should be in mind when we consider the FCNC constraints. 
Non-universality among squark masses
with different quantum numbers can bring various interesting results 
as suggested in \cite{ksy}.

In this paper we study the flavor structure of soft breaking parameters
in a certain type of effective supergravity theory which is derived from  
moduli/dilaton dominated supersymmetry breaking in superstring\cite{kl,bim}.
This recently proposed framework has some advantages.
We can use it without the knowledge of the origin of supersymmetry breaking.
Furthermore, within this framework it can give us rather detailed informations 
on the soft supersymmetry breaking parameters. As its result
it makes us to be able to extract their concrete flavor structure as seen
in the following discussion.
Taking account of these aspects, the study of soft supersymmetry breaking 
parameters based on this framework now seems worthy to be done more extensively
from various points of view.
The results we present in this paper are derived under strong assumptions
which, however, are expected to be applicable to rather wide class of 
superstring effective models. 

In the followings at first we briefly review the derivation of
soft breaking parameters in the case of moduli/dilaton dominated
supersymmetry breaking, which we take as the basis of our argument.
After that we present our basic assumptions.
Next under these assumptions we derive soft breaking parameters referring
to their flavor structure. Using these results, we give interesting sum rules
on their flavor structure and discuss the consequences derived from them.
Finally we study their alignment phenomena due to Yukawa couplings
and discuss their effects on the low energy physics briefly.
The last section is devoted to the summary. 

\section{Soft SUSY breaking parameters}
\subsection{General formulae}
We begin with the brief review of the general structure of soft breaking 
parameters in the case of moduli/dilaton dominated supersymmetry breaking.
Various works based on superstring theory and also general 
supergravity theory 
suggest that soft supersymmetry breaking parameters are generally 
non-universal\cite{il,kl,bim}.
Low energy effective supergravity theory is characterized 
in terms of the K\"ahler potential $K$, the superpotential $W$ and 
the gauge kinetic function $f_a$.
Each of these is a function of ordinary massless chiral matter 
superfields 
$\Psi^I$ and gauge singlet fields $\Phi^i$ called moduli\footnote{
Here we are using the terminology ``moduli'' in the generalized meaning.
A dilaton $S(\equiv \Phi^0)$ is included in $\Phi^i$ other than the
usual moduli $M^i(\equiv\Phi^i~(1\le i\le N))$. Throughout this paper
we will use this terminology as far as we do not state it.},
whose potential is perturbatively flat as far as supersymmetry 
is unbroken.

Usually it is assumed that the nonperturbative phenomena such as a 
gaugino condensation occur in a hidden sector.
After integrating out the fields relevant to these phenomena,
the K\"ahler potential and the superpotential are expanded by the low 
energy observable matter fields $\Psi^I$ as,
\begin{eqnarray}
&&K=\kappa^{-2}\hat K(\Phi,\bar \Phi)+
Z(\Phi,\bar \Phi)_{I{\bar J}}\Psi^I\bar \Psi^{\bar J}
+\left({1 \over 2}Y(\Phi,\bar \Phi)_{IJ}\Psi^I\Psi^J+{\rm h.c.}\right)
+\cdots, \\
&&W=\hat W(\Phi)+{1 \over 2}\tilde \mu(\Phi)_{IJ}\Psi^I\Psi^J+
{1 \over 3}\tilde h(\Phi)_{IJK}\Psi^I\Psi^J\Psi^K +\cdots,
\end{eqnarray}
where $\kappa^2=8\pi/M_{\rm pl}^2$.
The ellipses stand for the higher order terms in $\Psi^I$.
In eq.(2), $\hat W(\Phi)$ and $\tilde \mu(\Phi)_{IJ}$ are considered to be
induced by the nonperturbative effects in the hidden sector.
Using these functions the scalar potential $V$ can be written as
\cite{csgp},
\begin{equation}
\label{spot}
V=\kappa^{-2}e^G\left[G_\alpha(G^{-1})^{\alpha \bar \beta}G_{\bar \beta}
-3\kappa^{-2}\right]+({\rm D-term}),
\end{equation}
where $G=K+\kappa^{-2}\log \kappa^6 |W|^2$ and the indices $\alpha$ and
$\beta$ denote $\Psi^I$ as well as $\Phi^i$.
The gravitino mass $m_{3/2}$ which characterizes the scale of supersymmetry 
breaking is expressed as
\begin{equation}
m_{3/2}=\kappa^2e^{\hat K/2}|\hat W|.
\end{equation}
In order to get soft supersymmetry breaking terms in the low 
energy effective theory from eq.(\ref{spot}), we take the flat limit 
$M_{\rm pl}\rightarrow \infty$ preserving $m_{3/2}$ fixed.
Through this procedure we obtain the effective superpotential $W^{\rm eff}$
and soft supersymmetry breaking terms ${\cal L}_{\rm soft}$.

In the effective superpotential $W^{\rm eff}$, Yukawa couplings are 
rescaled as 
$h_{IJK}= e^{\hat K/2}\tilde h_{IJK}$ and a $\mu_{IJ}$ parameter is 
effectively 
expressed as
\begin{equation}
\label{mu}
\mu_{IJ} =e^{\hat K/2}\tilde \mu_{IJ} +m_{3/2}Y_{IJ} 
-F^{\bar j}\partial_{\bar j}Y_{IJ}.
\end{equation}
Soft breaking terms ${\cal L}_{\rm soft}$ corresponding to $W_{\rm eff}$
are defined by
\begin{equation}
{\cal L}_{\rm soft}=-\tilde m^2_{I\bar J}\psi^I\bar\psi^{\bar J}
-\left({1\over 3}A_{IJK}\psi^I\psi^J\psi^K +{1\over 2}B_{IJ}\psi^I\psi^J
+ {\rm h.c.}\right),
\end{equation}
where $\psi^I$ represents the scalar component of $\Psi^I$.
Each soft breaking parameter is expressed by using $K$ and $W$ as 
follows\cite{kl}\footnote{
It should be noted that these soft breaking parameters are not 
canonically normalized
because the kinetic term of $\psi^I$ is expressed as 
$Z_{I\bar J}\partial^\mu\psi^I\partial_\mu\bar\psi^{\bar J}$. },
\begin{eqnarray}
&&\tilde m_{I{\bar J}}^2=m_{3/2}^2Z_{I{\bar J}}
-F^i{\bar F}^{\bar j}\left[\partial_i\partial_{\bar j}Z_{I{\bar J}}
-\left(\partial_{\bar j}Z_{N{\bar J}}\right)Z^{N{\bar L}}
\left(\partial_iZ_{I{\bar L}}\right)\right]
+\kappa^2V_0Z_{I{\bar J}}, \label{smass} \\
&&A_{IJK}=F^i\left[\left(\partial_i +{1 \over 2}{\hat K}_i\right)h _{IJK}
-Z^{\bar ML}\partial_iZ_{\bar M(I} h_{JK)L}\right], \label{aterm} \\
&&B_{IJ}=F^i\left[\left(\partial_i +{1 \over 2}{\hat K}_i\right)\mu_{IJ}
-Z^{\bar ML}\partial_iZ_{\bar M(I} \mu_{J)L}\right]
-m_{3/2}\mu_{IJ} \nonumber \\
&&\qquad\qquad\qquad +\left[F^i\left(\partial_i 
+{1 \over 2}{\hat K}_i\right)F^{\bar j}
-2m_{3/2}F^{\bar j}\right]\partial_{\bar j}Y_{IJ}, \label{bterm}
\end{eqnarray}
where $F^i$ is an F-term of $\Phi^i$ and $\partial_i$ denotes
$\partial/\partial \Phi^i$.
$V_0$ is the cosmological constant expressed as
$V_0=\kappa^{-2}(F^i{\bar F}^{\bar j} \partial_i \partial_{\bar j}
\hat K-3m_{3/2}^2)$.
From these expressions we find that these soft breaking parameters are 
generally non-universal 
and their structure is dependent on the form of K\" ahler potential, 
especially, the functional form of $Z_{I\bar J}$.

The gaugino mass $M_a$ is derived through the following
formula\cite{csgp},
\begin{equation}
\label{mgauge}
M_a={1 \over 2}\left({\rm Re}~f_a\right)^{-1} F^j\partial_jf_a,
\end{equation}
where the subscript $a$ represents a corresponding gauge group.
In the superstring effective theory it is well known that $f_a=k_aS$ at the
tree level, where $k_a$ is the Kac-Moody level. 
It has the dependence on $M^i$ through the one-loop 
effects\cite{one}. This fact and eq.(10) bring an important result.
That is, if the dilaton contribution to the supersymmetry breaking
is large, the gaugino masses become large. On the other hand,
if gaugino masses are large enough, the difference among the squark masses 
disappears at the low energy region due to the radiative effects of heavy
gauginos. In this paper we are mainly interested in the non-universal soft
scalar masses. Thus in the following discussion we assume the small
gaugino masses implicitly.

Here it is necessary to make some comments on the application of these 
formulae to the MSSM.
The chiral superfields $\Psi^{I}$ represent quarks and leptons 
$Q^\alpha,$ $ \bar U^\alpha,$ $\bar D^\alpha,$
$L^\alpha$ and $\bar E^\alpha$ where $\alpha$ is a generation index. 
If only a pair of Higgs doublets are included, 
$\Psi^K$ in Yukawa couplings and the corresponding A-terms 
should be identified with $H_1$ and $H_2$. For this reason we will 
abbreviate the index $K$ of $A_{IJK}$ in eq.(8).
From the gauge invariance, allowed terms such as
$\mu_{IJ}\Psi^I\Psi^J$ in $W^{\rm eff}$ and  $Y_{IJ}\Psi^I\Psi^J$ in 
K\"ahler potential $K$
are only $\mu H_1H_2$ and $YH_1H_2$, respectively.
Taking account of these, the effective superpotential $W^{\rm eff}$ and
soft supersymmetry breaking terms ${\cal L}_{\rm soft}$ in the MSSM
can be written as
\begin{equation}
\label{sup}
W^{\rm eff}=h^U_{\alpha\beta}\bar U^\alpha H_2Q^\beta 
+h^D_{\alpha\beta}\bar D^\alpha H_1Q^\beta 
+h^E_{\alpha\beta}\bar E^\alpha H_1L^\beta +\mu H_1H_2,
\end{equation}
\begin{eqnarray}
\label{soft}
{\cal L}_{\rm soft}&=&-\sum_{I,J} z^{I\dag} \tilde m_{\bar IJ}^2z^J  
-\left(A^U_{\alpha\beta}\bar U^\alpha H_2Q^\beta 
+A^D_{\alpha\beta}\bar D^\alpha H_1D^\beta 
+A^E_{\alpha\beta}\bar E^\alpha H_1L^\beta 
\right.\nonumber \\
&& \qquad\qquad\qquad
+\left. BH_1H_2+\sum_a{1 \over 2}M_a\bar\lambda_a\lambda_a 
+ {\rm h.c.}\right).
\end{eqnarray}
The first term of eq.(12) represents the mass term of all scalar 
components $(z^I=Q^\alpha,$ $U^\alpha,$ $ D^\alpha,$ $ L^\alpha,$
$ E^\alpha,$ $ H_1, H_2)$ 
in the MSSM.
In the last term $\lambda_a$ are the gaugino fields for the gauge groups
specified by $a(a=3,2,1)$.

As seen from the general expressions of soft breaking parameters (7)$\sim$(9),
their structure is determined by the moduli dependence of $Z_{I\bar J}$ and
$W$.\footnote{
Only known exception is the dilaton dominated supersymmetry breaking.}
In order to apply these general results to eq.(12) and proceed further 
investigation of their flavor structure,
it is necessary to make the model more definite by introducing some 
assumptions.

\subsection{Assumptions}
Our basic assumptions are followings:

\noindent
(i)~we impose the simplest target space duality $SL(2,{\bf Z})$ invariance
\begin{equation}
M^i \rightarrow {a_iM^i-ib_i \over ic_iM^i+d_i} \qquad 
(a_id_i-b_ic_i=1 , ~a_i,b_i,c_i,d_i \in {\bf Z}),
\end{equation}
for each usual modulus $M^i$. 
Under this target space duality transformation (13) the chiral superfields 
$\Psi^I$ are assumed to be transformed as,
\begin{equation}
\Psi^I \rightarrow \left(ic_iM^i +d_i\right)^{n^i_I}\Psi^I,
\end{equation}
where $n^i_I$ is called the modular weight and takes a suitable negative 
value\cite{il}.
This requirement also causes the invariance under the 
following K\"ahler transformation,
\begin{eqnarray}
&&K \rightarrow K +f(M^i)+\bar f(\bar M^{\bar i}),\\
&&W \rightarrow e^{-f(M^i)}W,
\end{eqnarray}
(ii)~K\"ahler metric and then the kinetic terms of chiral superfields
are flavor diagonal
\begin{equation}
Z_{I\bar J}=Z_I\delta_{IJ},
\end{equation}
(iii)~coefficient functions $\tilde h_{IJK}, \tilde\mu_{IJ}, Y_{IJ}$ of
the superpotential and K\"ahler potential are independent of the moduli 
fields whose F-terms 
contribute the supersymmetry breaking, that is,
\begin{equation} 
\partial_i\tilde h_{IJK}=\partial_i\tilde\mu_{IJ}=\partial_i Y_{IJ}=0
\qquad {\rm for}~F^i \not=0.
\end{equation}
The second assumption is satisfied in the almost all known superstring models
as suggested in \cite{bim}.
The third one is a pure assumption at this level. 
Under this assumption the superpotential $W$, except for a $\hat W$ part,
can depend only on moduli which
do not contribute the supersymmetry breaking and as its result, for example,
Yukawa couplings $\tilde h_{IJK}$ can be dynamical variables at the low
energy region as discussed in ref.\cite{bd}.
These assumptions are rather strong ones but they may be expected to be 
satisfied in many known superstring models.
Moreover, they can induce very interesting features to the soft breaking
parameters as seen in the following parts.

In order to parametrize the direction of supersymmetry breaking in the moduli
space,
we introduce the parameters $\Theta_i$ which correspond to the generalized
Goldstino angles in the moduli space\cite{bim}.
They are defined as
\begin{equation}
F^i\sqrt{\hat K_{i\bar j}}=\sqrt 3Cm_{3/2}\Theta_i,
\qquad \sum_{i=0}^N\Theta^2_i =1,
\end{equation}
where we take the $\kappa=1$ unit. $N$ is the number of usual moduli 
$M^i$ in the model.
A constant $C$ satisfies $V_0=3\kappa^{-2}(|C|^2-1)m_{3/2}^2$.
In the followings we assume $C=1$ and then $V_0=0$.
The introduction of these parameters makes it possible to discuss the soft 
breaking parameters without asking the origin of supersymmetry breaking.

\subsection{Sum rules for soft scalar masses}
In the superstring models studied by now, $\hat K$ can be generally written as
\begin{equation}
\hat K=-\sum_{i=0}^N\ln\left(\Phi^i +\bar \Phi^{\bar i}\right).
\end{equation}
On the other hand, 
if we apply the assumptions (i) and (ii) to $Z_{I\bar J}$ in the K\"ahler
potential $K$, we can constrain the functional form of $Z_I$ as
\begin{equation}
Z_I=\prod_{i=1}^N (M^i +\bar M^{\bar i})^{n^i_I}.
\end{equation}
Using these facts in eqs.(7)$\sim$(9) and normalizing them canonically,
we can write down the soft breaking 
parameters as\footnote{It should be noted that these are the tree level
results. However, the introduction of string one-loop effects will not change
the qualitative features discussed here. }
\begin{eqnarray}
&&\tilde m_I^2=m_{3/2}^2\left( 1 +3\sum_{i=1}^N\Theta_i^2n_I^i \right), \\
&&A_{IJ}=-\sqrt 3m_{3/2}h_{IJ}\sum_{i=0}^N\Theta_i\left(n_I^i+n_J^i
+n_{H_{1,2}}^i+1 \right), \\ 
&&B=-m_{3/2}\mu\left[\sqrt 3\sum_{i=0}^N\Theta_i
\left(n_{H_1}^i+n_{H_2}^i +1 \right)+1\right],
\end{eqnarray}
where the indices $I$ and $J$ represent the flavors $Q_\alpha, \bar U_\alpha,
\bar D_\alpha, L_\alpha$ and $\bar E_\alpha (\alpha =1 \sim 3)$.
In these formulae $n_I^0=0$ should be understood since we do not consider 
the transformations such as (13) and (14) for a dilaton.

Taking account of the functional form of $\hat K$ in eq.(20), 
the assumption (i) requires $f(M^i)=-\ln(ic_iM^i+d_i)$ in eq.(15).
As a result, eq.(16) shows that the superpotential $W$ and then its 
coefficient functions $\tilde h_{IJ}$ and $\tilde\mu_{IJ}$ are
transformed as the modular forms under the duality transformation of 
moduli fields $M^i$,
\begin{eqnarray}
&&W \rightarrow \prod_{i=1}^N\left( ic_iM^i +d_i\right)^{-1}W, \\
&&\tilde h_{IJ} \rightarrow \prod_{i=1}^N\left( 
ic_iM^i +d_i\right)^{(-n_I^i-n_J^i-n_{H_{1,2}}^i-1)} \tilde h_{IJ}, \\
&&\tilde \mu_{IJ} \rightarrow \prod_{i=1}^N\left( 
ic_iM^i +d_i\right)^{(-n_{H_1}^i-n_{H_2}^i-1)}\tilde \mu_{IJ}.
\end{eqnarray}
The assumption (iii) requires that the modular weights of $\tilde h_{IJ}$ and 
$\tilde\mu_{IJ}$ for moduli $M^i$ which are relevant to the supersymmetry
breaking $(F_i\not=0)$ should be equal to 
zero. This results in the following relations
\begin{eqnarray}
&&n_I^i+n_J^i+n_{H_{1,2}}^i+1=0, \\
&&n_{H_1}^i+n_{H_2}^i+1=0,
\end{eqnarray}
for each $i(\not=0)$. 
Flavor indices $I$ and $J$ in eq.(28) should be taken as the ones composing 
each Yukawa
coupling in eq.(11). After substituting these relations into 
eqs.(22)$\sim$(24), we obtain the formulae 
for soft breaking parameters,
\begin{eqnarray}
&&\tilde m_{Q_\alpha}^2=m_{3/2}^2\left( 1 
+3\sum_{i=1}^N\Theta_i^2n_{Q_\alpha}^i \right), \\
&&\tilde m_{\bar U_\alpha}^2=m_{3/2}^2\left( 1 +3\sum_{i=1}^N\Theta_i^2
(-n_{Q_\alpha}^i-n_{H_2}^i-1) \right), \\
&&\tilde m^2_{\bar D_\alpha}=m_{3/2}^2\left( 1 +3\sum_{i=1}^N\Theta_i^2
(-n_{Q_\alpha}^i-n_{H_1}^i-1) \right), \\
&&\tilde m_{L_\alpha}^2=m_{3/2}^2\left( 1 +3\sum_{i=1}^N\Theta_i^2
(-n_{\bar E_\alpha}^i-n_{H_2}^i-1) \right), \\
&&\tilde m_{\bar E_\alpha}^2=m_{3/2}^2\left( 1 +3\sum_{i=1}^N\Theta_i^2
n_{\bar E_\alpha}^i \right),\\
&&A_{IJ}=-\sqrt 3m_{3/2}h_{IJ}\Theta_0, \\ 
&&B=-\left(\sqrt 3\Theta_0+1\right)m_{3/2}\mu.
\end{eqnarray}

The results for parameters $A_{IJ}$ and $B$ are similar to the ones 
obtained in the case
of dilaton dominated supersymmetry breaking. 
These features are brought about
by the assumption (iii). Although soft scalar masses are flavor diagonal,
those values are non-universal unlike the dilaton dominated case.
Their non-universality is
determined by the modular weights of the relevant fields,
which are completely dependent on the models.
This feature seems to make it difficult to practice the model 
independent study of the flavor 
structure of soft scalar masses in the present model.
However, it is remarkable that we can easily 
extract some informations on the flavor structure of soft scalar masses
in the model independent way by constructing the sum rules from 
these formulae.
We present here two sum rules at $M_{\rm pl}$,\footnote{The similar sum 
rules are derived in 
\cite{bims}. However, the flavor structure is not explicitly discussed there.}
\begin{eqnarray}
&&\tilde m_{Q_\alpha}^2+\tilde m_{\bar D_\alpha}^2=\tilde m_{L_\alpha}^2
+\tilde m_{\bar E_\alpha}^2, \\
&&2\tilde m_{Q_\alpha}^2+\tilde m_{\bar U_\alpha}^2+\tilde m_{\bar D_\alpha}^2
=m_{3/2}^2\left(4-3\sum_{i=1}^N\Theta_i^2 \right),
\end{eqnarray}
where for these derivations we used the above constraints (28) and (29) 
on the modular weights.
These sum rules for the flavors are satisfied in each generation($\alpha=1
\sim 3$).

Although these sum rules become trivial in the universal case,
they can give us the useful informations on 
the flavor structure of soft scalar masses
in the non-universal situation.
Especially, the latter sum rule (38) gives us very interesting insights of 
the flavor structure in the quark sector.

At first it shows that we can not impose the relation 
$\tilde m_{f_\alpha}^2 < \tilde m_{f_\beta}^2$ ($\alpha <\beta$) 
on all flavors
$f=Q, \bar U$ and $\bar D$, simultaneously.
This means that soft scalar masses at least for one flavor 
in the squark sector must decrease
according as the generation number increases 
though scalar masses in  other flavor sectors increase with
the generation.
For example, we assume that soft scalar masses of the first two generation
with the same charges degenerate each other and
\begin{equation}
\tilde m_{Q_1}^2=\tilde m_{Q_2}^2 < \tilde m_{Q_3}^2,\qquad
\tilde m_{\bar D_1}^2=\tilde m_{\bar D_2}^2 < \tilde m_{\bar D_3}^2.
\end{equation}
In this situation, from eq.(38) we have
\begin{equation}
\tilde m_{\bar U_1}^2=\tilde m_{\bar U_2}^2 > \tilde m_{\bar U_3}^2.
\end{equation}
This feature can bring nontrivial results at the weak scale through
dynamical effects of the low energy region as 
seen in the next section.

Furthermore, since the sum of squark masses is constrained to be constant 
independently of the generation, if the soft scalar mass for one flavor, 
for example, $\tilde m_{Q_\alpha}^2$
becomes larger, soft scalar masses $\tilde m_{\bar U_\alpha}^2$
and $\tilde m_{\bar D_\alpha}^2$ must be smaller in comparison with  
$\tilde m_{Q_\alpha}^2$.\footnote{It can be shown that this kind 
of hierarchical 
soft scalar masses can be realized if we consider three moduli case in 
orbifold models\cite{ksyy}.}
In such a case eq.(37) shows that in the present model all left-handed soft
scalar masses can be larger than all right-handed ones,
\begin{equation}
\tilde m^2_{Q_\alpha},~\tilde m_{L_\alpha}^2 \gg \tilde m^2_{\bar U_\alpha},
~\tilde m^2_{\bar D_\alpha},~\tilde m^2_{\bar E_\alpha}.
\end{equation}
Soft scalar masses with this feature have been shown to bring various 
interesting implications in the phenomenology of the MSSM\cite{ksy}.

In the next section we will study the low energy implication of 
these structures.
In this study we will consider the alignment of soft scalar masses 
in the flavor space which has recently been proposed in ref.\cite{dgt}.

\section{Alignment of soft scalar masses}
\subsection{Flavor symmetry}
We start this section with the discussion on the flavor symmetry of the system
defined by the K\"ahler potential (1) and the superpotential (2).
What we refer to here as the flavor symmetry is the invariance under the 
following transformation,
\begin{equation}
\Psi^{f_\alpha} \rightarrow S^{(f)}_{\beta\alpha}\Psi^{f_\alpha},
\end{equation}
where $\Psi^{f_\alpha}$ stands for $Q_\alpha, U_\alpha, D_\alpha,
L_\alpha$ and $E_\alpha~(\alpha =1\sim 3)$ and $S^{(f)}$ is an 
element of $U(3)$. Thus the full flavor symmetry of our present model is 
$U(3)^5$.
Needless to say, Yukawa couplings in the superpotential break this symmetry. 
Even if we switch off these couplings, this symmetry is also broken by 
the kinetic terms unless $Z_{I\bar J}$ is proportional to the unit 
matrix.  This condition for the kinetic terms is not generally
satisfied in the present model.
However, there may be an alternative possibility.
If there are some relations between the moduli space and the flavor space,
the moduli dependence of $Z_{I\bar J}$ may restore the symmetry.
That is, it may be expected that $Z_{I\bar J}(\equiv 
Z^{(f)}_{\alpha\bar\beta})$ is also transformed simultaneously 
under the transformation (42) as
\begin{equation}
Z^{(f)}_{\alpha\bar\beta} \rightarrow S^{(f)}_{\gamma\alpha}
Z^{(f)}_{\alpha\bar\beta}S^{(f)\dag}_{\bar\beta\bar\delta},
\end{equation}
and then this can be the symmetry of the system.

If such a situation is realized in the present supersymmetry breaking
scenario, VEVs of moduli F-terms and then the soft supersymmetry breaking
parameters will cause the breaking of this flavor symmetry as well as
supersymmetry.
In this breaking process the dynamical degrees of freedom corresponding 
to the Goldstone modes of this spontaneously broken flavor group will 
remain undetermined unless explicit breakings exist.
However, there are Yukawa couplings in the real world.
As its result the phenomenon known as the alignment occurs to fix the 
remaining undetermined degrees of freedom in the soft 
breaking parameters at the 
low energy region. 
Recently, this possibility has been
suggested in the general framework of supersymmetric models 
assuming the existence of the corresponding
flavor symmetry\cite{dgt}.
In the remaining part we study the alignment of soft breaking parameters 
in our model.
In this study we assume the transformation property (43) in the moduli 
sector as the starting point of our argument.\footnote{
It should be noted that this assumption is crucial for the realization of 
the alignment of soft breaking parameters in the flavor space due to 
Yukawa couplings.
Although the contributions from the Goldstone boson loops to the effective
potential are sufficiently suppresssed by the symmetry breaking scale
$\Lambda$ in the case of explicit breakings due to Yukawa couplings,
their contributions through the breakings due to $Z_{I\bar J}$ in the kinetic 
terms can not be suppressed by $\Lambda$ and make the scenario in \cite{dgt}
ineffective. From this reason we will adopt this assumption, although
the existence of such a proprety has not been known in the superstring models
by now.}

\subsection{Alignment in the flavor space}
As seen from eqs.(30)$\sim$(36), soft scalar masses and A-terms are produced
in the flavor diagonal form in the present model.
In the limit that Yukawa couplings are negligible, 
however, the degrees of freedom corresponding to the rotation 
$\left(SU(3)/U(1)^2\right)^5$ in the flavor space remain undetermined 
under the above assumption on the flavor symmetry.
If we represent these degrees of freedom
 with $3\times 3$ matrices $S^{(f)}$ in the basis where
Yukawa couplings $h^F~(F=U,D,E)$ are diagonal, soft breaking 
parameters $\tilde m_f^2$ and $A^F(\equiv A_{IJ})$ 
can be written as\footnote{It should be noted that $S^{(f)}$ for the flavor
$\bar U, \bar D$ and $\bar E$ are defined as the ones for the right-handed
$U, D$ and $E$ in eq.(42).}
\begin{equation}
\tilde m^2_f=S^{(f)\dag}\Sigma^{(f)}S^{(f)}, \qquad 
A^F=S^{(f^\prime)\dag}\Delta^FS^{(f)}.
\end{equation}
From the definition of the Yukawa couplings (11) the index $f^\prime$
in the representation of $A^F$ stands for the flavor $f^\prime = U, D, E$
which can compose the Yukawa couplings with the flavor $f=Q, L$.
Both of $\Sigma^{(f)}$ and $\Delta^F$ correspond to the ones derived 
in the previous section
and they are defined at the scale $\Lambda=M_{\rm pl}$.
The Goldstone degrees of freedom
$S^{(f)}$ are determined through the physics below the scale $\Lambda$.
To study this determination process we adopt Wilsonian approach 
to the low energy effective theory.
Following ref.\cite{dgt},
the low energy effective potential which can be derived by such a prescription
is
\begin{equation}
V_{\rm eff}=V_\Lambda +{\Lambda^2 \over 32\pi^2}Str\left(
{\cal M}^2 -{1 \over 32\pi^2}\beta^{(1)}_{\cal M}+ \cdots \right),
\end{equation}
where $V_\Lambda$ is an $S^{(f)}$ independent part and ${\cal M}$ represents
a mass matrix of the fields in the theory at the scale $\Lambda$.
The ellipses stand for the higher order correction terms which are irrelevant 
in the present approximation. 
One-loop $\beta$-functions $\beta^{(1)}_{\cal M}$ for the masses of the 
relevant scalar fields are given in ref.\cite{mv},
\begin{eqnarray}
&&Tr\beta^{(1)}_{\tilde m_Q^2}=
Tr\left[2\left( h^{U\dag} h^U+h^{D\dag} h^D\right) \tilde m_Q^2
+2h^{U\dag} \tilde m_{\bar U}^2h^U +2h^{D\dag} 
\tilde m_{\bar D}^2h^D\right. \nonumber \\ 
&&\qquad\qquad\qquad 
\left.+2A^{U\dag}A^U+2A^{D\dag}A^D \right]+\cdots , \nonumber \\ 
&&Tr\beta^{(1)}_{\tilde m_L^2}=
Tr\left[\tilde m^2_Lh^{E\dag}h^E+2h^{E\dag} \tilde m_{\bar E}^2h^E
+h^{E\dag} h^E \tilde m_L^2+2A^{E\dag}A^E \right]+\cdots , \nonumber \\ 
&&Tr\beta^{(1)}_{\tilde m_{\bar U}^2}=
Tr\left[2\tilde m^2_{\bar U}h^Uh^{U\dag}+4h^U \tilde m_Q^2h^{U\dag}
+2h^Uh^{U\dag} \tilde m_{\bar U}^2+4A^UA^{U\dag} \right]+\cdots , \nonumber \\ 
&&Tr\beta^{(1)}_{\tilde m_{\bar D}^2}=
Tr\left[2\tilde m^2_{\bar D}h^Dh^{D\dag}+4h^D \tilde m_Q^2h^{D\dag}
+2h^Dh^{D\dag} \tilde m_{\bar D}^2+4A^DA^{D\dag} \right]+\cdots , \nonumber \\ 
&&Tr\beta^{(1)}_{\tilde m_{\bar E}^2}=
Tr\left[2\tilde m^2_{\bar E}h^Eh^{E\dag}
+4h^E \tilde m_L^2h^{E\dag} +2h^Eh^{E\dag} \tilde m_{\bar E}^2
+4A^EA^{E\dag} \right]+\cdots , \nonumber \\ 
&&Tr\beta^{(1)}_{\tilde m_{H_2}^2}=
6Tr\left[\tilde m^2_Qh^{U\dag}h^U
+h^{U\dag} \tilde m_{\bar U}^2h^U+A^{U\dag}A^U \right]+\cdots , \nonumber \\ 
&&Tr\beta^{(1)}_{\tilde m_{H_1}^2}=
Tr\left[ 6\tilde m^2_Qh^{D\dag}h^D
+2\tilde m_L^2h^{E\dag} h^E+6h^{D\dag} \tilde m_{\bar D}^2h^D
+2h^{E\dag} \tilde m_{\bar E}^2h^E\right. \nonumber\\
&&\qquad\qquad\qquad\left. +6A^{D\dag}A^D+2A^{E\dag}A^E \right]+\cdots , 
\end{eqnarray}
where the ellipses stand for the $S^{(f)}$ independent contributions.
Following (44), the soft scalar masses in these formulae are expressed 
in the Yukawa couplings diagonal basis by using the nonlinearly 
realized Goldstone modes $S^{(f)}$ as follows,
\begin{eqnarray}
&&\tilde m_{Q_U}^2 =S^{(Q)\dag} \Sigma^{(Q)} S^{(Q)}, \nonumber \\
&&\tilde m_{Q_D}^2 ={\cal K}^{\dag} S^{(Q)\dag}\Sigma^{(Q)} S^{(Q)} 
{\cal K}, \nonumber \\
&&\tilde m_{\bar U}^2 =S^{(U)\dag}\Sigma^{(U)} S^{(U)}, 
\nonumber \\
&&\tilde m_{\bar D}^2 =S^{(D)\dag}\Sigma^{(D)}S^{(D)},
\end{eqnarray}
where ${\cal K}$ is the Kobayashi-Maskawa matrix.
Using these, we can write down the $S^{(f)}$ dependent part of this 
effective potential,
\begin{eqnarray}
&&V(S^{(f)})=-{\Lambda^2\over (32\pi^2)^2}
Tr\left[12S^{(Q)\dag}\Sigma^{(Q)} S^{(Q)}\left(h^{U\dag} h^U
+{\cal K} h^{D\dag} h^D{\cal K}^{\dag}
\right)\right. \nonumber \\
&&\qquad\qquad\left. +8S^{(L)\dag}\Sigma^{(L)}S^{(L)} h^{E\dag} h^E
+12S^{(U)\dag}\Sigma^{(U)}S^{(U)}h^Uh^{U\dag}\right. \nonumber \\
&&\qquad\qquad\left. +12S^{(D)\dag}\Sigma^{(D)}S^{(D)}h^Dh^{D\dag} 
+8S^{(E)\dag}\Sigma^{(E)}S^{(E)}h^Eh^{E\dag}\right]. 
\end{eqnarray}
Here it should be noted that the $A^F$ contribution to $V_{\rm eff}$ disappears
in eq.(48) because $A^F$ appears as the unitary invariant form 
in the effective potential
in this approximation level.
The flavor structure of soft breaking parameters at the weak scale 
is determined through eq.(44) by using $S^{(f)}$ which minimizes 
this effective potential.

As easily proved for the diagonal matrices $X$ and $Y$, 
~$Tr(XS^{(f)}YS^{(f)\dag})$
is maximized when $S^{(f)}$ is aligned so as for their eigenvalues $X_i$ and
$Y_i$ to be ordered in such a way that $X_i\le X_j$ and $Y_i \le Y_j$ are
simultaneously realized for the diagonal component labels $i<j$.
We are considering the effective potential (48) in the basis where the
Yukawa coupling matrices are diagonal and also their eigenvalues are 
ordered in such a way that their values increase.
So the potential minimization makes $S^{(f)}~(f=L, U, D, E)$  be a unit 
matrix if the eigenvalues of $\Sigma^{(f)}$ are ordered in a suitable way. 
Thus soft scalar masses for these flavor at the weak scale
are completely aligned to Yukawa couplings in the flavor space.
In the $f=Q$ sector the situation is different.
Because of the existence of the Kobayashi-Maskawa matrix ${\cal K}$,
$S^{(Q)}$ deviates from the unit matrix even if the eigenvalues of 
$\Sigma^{(Q)}$ are ordered according to their magnitude.
From the investigation of the effective potential it is easily found that
$S^{(Q)}$ should be determined so that
$S^{(Q)}(h^{U\dag} h^U+{\cal K}h^{D\dag} h^D{\cal K}^{\dag})S^{(Q)\dag}$ is
 propotinal to $\Sigma^{(Q)}$.
As found from eq.(44), these facts show that A-parameters in the 
squark sector are not generally propotional to the Yukawa couplings,
although $A^L$ in the lepton sector is propotional to $h^L$.

Here we should remind again that in our model
the following condition for $\Sigma^{(f)}$ is satisfied independently
of the generation,
\begin{equation}
2\Sigma^{(Q_\alpha)}+\Sigma^{(U_\alpha)}+\Sigma^{(D_\alpha)}=
{\rm const.}
\end{equation}
and then the eigenvalues of
$\Sigma^{(f)}$ are not necessarily ordered in such a way that
their magnitude increases for all $f$ as mentioned before. 
In that case the minimization of the effective potential 
determines $S^{(f)}(\not=1)$ so as to exchange the ordering of the 
eigenvalues according to their magnitude. 
The existence of the Kabayashi-Maskawa matrix ${\cal K}$
makes this situation more subtle.
These make the flavor structure of A-parameters somehow complex
at least in the squark sector.

\subsection{Squark sector}
We study the squark sector in more detail.
The squark mass matrices in u-squark and d-squark sectors can be written as,
\begin{equation}
\left( \begin{array}{cc}
\vert m_U\vert^2 +\tilde m_{Q_U}^2
+m_Z^2\cos2\beta({1\over 2}-{2\over 3}\sin^2\theta_W) & 
A^U\langle H_2\rangle +m_U\mu^\ast \cot\beta \\
A^{U\dag}\langle H_2 \rangle^\ast + m_U^{\dag}\mu\cot\beta & 
\vert m_U\vert^2 +\tilde m_{\bar U}^2
+{2\over 3}m_Z^2\cos2\beta \sin^2\theta_W 
\end{array}\right),
\end{equation}
\begin{equation}
\left( \begin{array}{cc}
\vert m_D\vert^2 +\tilde m_{Q_D}^2
-m_Z^2\cos2\beta({1\over 2}-{1\over 3}\sin^2\theta_W) & 
A^D\langle H_1\rangle +m_D\mu^\ast \tan\beta \\
A^{D\dag}\langle H_1\rangle^\ast + m_D^{\dag}\mu\tan\beta 
& \vert m_D\vert^2 +\tilde m_{\bar D}^2
-{1\over 3}m_Z^2\cos2\beta \sin^2\theta_W 
\end{array}\right),
\end{equation}
where $m_f$ and $\tilde m_{f}$ are masses of the 
$f$-quark, the corresponding left-handed squark and the right-handed 
squark mass matrices, respectively.
These squark mass matrices can be explicitly determined in our model.
We show this by using a typical example.

As discussed above, no matter how eigenvalues of $\Sigma^{(f)}$ are ordered in
the squark sector, under the condition of (49) we obtain the nontrivial
results for $S^{(Q)}, S^{(U)}$ and $S^{(D)}$ as far as the $\Sigma^{(f)}$ 
has the nondegenerate eigenvalues.
We are interested in the nontrivial case where
the eigenvalues of $\Sigma^{(f)}$ are not ordered according to their 
magnitude.
As such a typical example, we take
\begin{equation}
\Sigma^{(Q)}=\left( \begin{array}{ccc}
m_1^2&0&0\\ 0&m_1^2&0 \\ 0&0&m_2^2 \\ \end{array}\right), \qquad
\Sigma^{(U)}=\Sigma^{(D)}=\left( \begin{array}{ccc}
m_3^2&0&0\\ 0&m_3^2&0 \\ 0&0&m_4^2 \\ \end{array}\right), 
\end{equation} 
where $m_1^2 >m_2^2$ and $m_4^2 > m_3^2$ following eq.(49).
For the $\bar U$ and $\bar D$ sectors we get $S^{(U)}=S^{(D)}=1$ trivially.
If ${\cal K}=1$,  for the $Q$ sector we obtain
\begin{equation}
S^{(Q)}=\left( \begin{array}{ccc}
0&1&0\\ 0&0&-1 \\ 1&0&0 \\ \end{array}\right). 
\end{equation} 
For this $S^{(Q)}$, $S^{(Q)\dag}\Sigma^{(Q)}S^{(Q)}$ becomes diagonal 
with increasingly ordered eigenvalues in the same way as 
the case of $\bar U$ and 
$\bar D$ sectors. 
If ${\cal K}\not=1$, however,  $S^{(Q)\dag}\Sigma^{(Q)}S^{(Q)}$ can 
not be diagonal
anymore. After some algbra for the minimization of the effective potential,
we obtain
\begin{equation}
S^{(Q)}=\left( \begin{array}{ccc}
0&1&X\\ Y&X&-1 \\ 1&0&Y \\ \end{array}\right),
\end{equation} 
where $X\sim -V_{ts}m_b^2/m_t^2$ and $Y\sim -V_{td}m_b^2/m_t^2$.
This results in
\begin{equation}
S^{(Q)\dag} \Sigma^{(Q)}S^{(Q)}\sim \left( \begin{array}{ccc}
Y^2m_1^2+m_2^2&XYm_1^2&-Ym_1^2+Ym_2^2\\ 
XYm_1^2&m_1^2+X^2m_1^2&0 \\ 
-Ym_1^2+Ym_2^2&0&X^2m_1^2+m_1^2+Y^2m_2^2 \\ \end{array}\right).
\end{equation} 
This shows that the effect of ${\cal K}\not=1$ is very small and 
$S^{(Q)\dag}\Sigma^{(Q)} S^{(Q)}$ can be regarded diagonal in the 
good approximation even in the $Q$ sector.
However, it should be noted that the degeneracy between the first
and second generation squarks in the $Q$ sector at $M_{\rm pl}$ is
lost at the low energy region.  

If we use eq.(44), we can explicitly calculate $A^U$ and $A^D$ in this 
example as follows,
\begin{equation}
A^U=\Delta S^{(U)\dag}h^US^{(Q)}\sim 
{e\Delta \over \sqrt 2 m_W\sin\beta\sin\theta_W}\left( \begin{array}{ccc}
0&m_u &Xm_u \\ Ym_c &Xm_c&-m_c \\ m_t&0&Ym_t \\ \end{array}\right),
\end{equation}
\begin{equation}
A^D=\Delta S^{(D)\dag}h^DS^{(Q)}\sim 
{e\Delta \over \sqrt 2 m_W\cos\beta\sin\theta_W}\left( \begin{array}{ccc}
0&m_d &Xm_d \\ Ym_s &Xm_s&-m_s \\ m_b&0&Ym_b \\ \end{array}\right),
\end{equation}
where $\Delta$ is defined as $\Delta^F=\Delta h^F$.
This result shows that $A^U$ and $A^D$ can have rather large off-diagonal
elements unless $S^{(Q)}=S^{(U)}=S^{(D)}=1$, which is realized only
when the masses of three generation squarks with the same quantum 
numbers degenerate.
However, the non-universal soft scalar masses such as (52) appear only 
in the case of 
moduli dominated supersymmetry breaking ($\displaystyle 
\sum_{i=1}^N\Theta_i^2\sim 1$ and $\Theta_0^2\ll 1$) and then
$\Delta^F \ll m_{3/2}$.
Thus the components of off-diagonal blocks in squark mass matrices
(50) and (51) are small compared with the elements of the diagonal blocks.
In our model the left and right mixing squark masses generally bring
no serious problems to the FCNC.

From this example we learn that if the flavor alignment occurs 
in the squark masses, the good degeneracy between the masses of 
the first and second generation squarks at
$M_{\rm pl}$ can be lost at the low energy region.
This shows that the good degeneracy at $M_{\rm pl}$ among the masses
of three generation squarks with the same quantum numbers may be 
necessary to satisfy the FCNC constraints.

The consideration like this can give us more insights for the soft scalar 
masses. 
Finally we present such a typical example in the nondegenerate case.
Taking account of the alignment effects discussed above,
if we assume
\begin{equation}
\tilde m_{Q_1}^2 \simeq \tilde m_{Q_2}^2 < \tilde m_{Q_3}^2, \qquad
\tilde m_{\bar D_1}^2 \simeq \tilde m_{\bar D_2}^2 < \tilde m_{\bar D_3}^2
\end{equation}
at the scale $\Lambda$, these relations are preserved at the weak scale.
This situation is favorable to satisfy the FCNC constraints.
On the other hand, the sum rule at $\Lambda$ and the alignment effects predict
\begin{equation}
\tilde m_{\bar U_1}^2 < \tilde m_{\bar U_2}^2 \simeq \tilde m_{\bar U_3}^2
\end{equation}
at the low energy region.
This suggests that in the present model it is difficult to make only
$\tilde m_{\bar U_3}^2$ light enough in comparison with other right handed
$U$ sector squarks.

\section{Summary}
We studied the structure of soft supersymmetry breaking parameters in the
effective theory derived from moduli/dilaton dominated supersymmetry
breaking. In particular, we focussed our attension on the flavor structure 
of soft scalar masses. 
Under certain assumptions on the moduli dependence of K\"ahler potential
and superpotential we obtained the sum rules for the
soft scalar masses, which gave the interesting relations among different 
flavors in the model independent way. 
We showed that  we could extract some typical 
features from these sum rules when scalar masses were non-universal.

We also applyed these results as the initial conditions at $M_{\rm pl}$
to the recently proposed 
alignment scenario for the soft scalar masses in the flavor space.
In this discussion we pointed out that the degeneracy of 
the masses among three generation 
squarks with the same quantum numbers might be required at $M_{\rm pl}$ to
satisfy the FCNC constraints. Only the good degeneracy 
between the first and second generation squark masses at $M_{\rm pl}$ may
not be necessarily sufficient to avoid the excessive FCNC.

The consequences induced from the non-universality of soft supersymmetry 
breaking parameters have been studied from various view points by now.
Although FCNC constraints require the degeneracy of 
squark masses among the species with the same quantum numbers, it does not
require the universality more than that.
It seems to be necessary to take account of this important point
when we study the flavor structure of soft breaking 
parameters.
This view point may open the new possibility in the study of the 
supersymmetric theory.
  
\vspace{7mm}
\noindent
%{\Large\bf Acknowledgement}\\ 
The author is grateful for the hospitality of TH-Division of CERN
where this work was done.
This work is supported in part by a Grant-in-Aid for Scientific 
Research from the Ministry of Education, Science and Culture
(\#05640337 and \#08640362).

\newpage
%%%%%%%%%%%%%%%%%%%%%%%%%%%% References %%%%%%%%%%%%%%%%%%%%%%%%%%%%%%%%%%%


\begin{thebibliography}{99}
\bibitem{n}For a review, see for example, H.-P.~Nilles, Phys. Rep.
{\bf C110}, 1 (1984), and references therein.

\bibitem{fcnc}F.~Gabbiani and A.~Masiero, Nucl. Phys. {\bf B322}, 235 (1989);

J.~Hagelin, S.~Kelley and T.~Tanaka, Nucl. Phys. {\bf B415}, 293 (1994), 
and references therein. 

\bibitem{edmn}Y.~Kizukuri and N.~Oshimo, Phys. Rev. {\bf D46}, 3025 (1992),
 and references therein.

\bibitem{dks}L.~J.~Hall, J.~Lykken and S.~Weinberg, Phys. Rev. 
{\bf D27}, 2359 (1983);

M.~Dine, A.~Kagan and S.~Samuel, Phys. Lett. 
{\bf B243}, 250 (1990);

M.~Dine and A.~E.~Nelson, Phys. Rev. {\bf D48}, 1277 (1993);
 
M.~Dine, R.~Robert and A.~Kagan, Phys. Rev. {\bf D48}, 4269 (1993); 

Y.~Nir and N.~Seiberg, Phys. Lett. {\bf B309}, 337 (1993);

L.~J.~Hall and H.~Murayama, Phys. Rev. Lett. 
{\bf 75}, 3985 (1995).

\bibitem{il}L.~E.~Ib\'a\~nez and D.~L\"ust, Nucl. Phys. 
{\bf B382}, 305 (1992).

\bibitem{kl}V.~S.~Kaplunovsky and J.~Louis, 
Phys. Lett. {\bf B306}, 269 (1993).

\bibitem{bim}A.~Brignole, L.~E~.Ib\'{a}\~{n}ez and C.~Mu\~{n}oz,
Nucl. Phys. {\bf B422}, 125 (1994).

\bibitem{mn}A.~Lleyda and C.~Mu\~noz, Phys. Lett. {\bf B317}, 82 (1993);

N.~Polonsky and A.~Pomarol, Phys. Rev. Lett. {\bf 73}, 2292 (1994);

D.~Matalliotakis and H.~P.~Nilles, Nucl. Phys. {\bf B435}, 115 (1995);

M.~Olechowski and S.~Pokorski, Phys. Lett. {\bf B344}, 201 (1995);

D.~Choudhury, F.~Eberlein, A.~K\"onig, J.~Louis and S.~Pokorski,
Phys. Lett. {\bf B342}, 180 (1995).

\bibitem{ksy}T.~Kobayashi, D.~Suematsu and Y.~Yamagishi, Phys. Lett.
{\bf B329}, 27 (1994);

T.~Kobayashi, M.~Konmura, D.~Suematsu, K.~Yamada and T.~Kobayashi,
Prog. Theor. Phys. {\bf 94}, 417 (1995).

\bibitem{csgp}E.~Cremmer, S.~Ferrara, L.~Girardello and A.~Van Proeyen,
Nucl. Phys. {\bf B212}, 413 (1983).

\bibitem{one}L.~E.~Ib\'a\~nez and H.~P.~Nilles, Phys. Lett. {\bf B169},
345 (1986);

L.~Dixon, V.~Kaplunovsky and J.~Louis, Nucl. Phys. {\bf B355}, 649 (1991).

\bibitem{bd}Y.~Nambu, preprint EFI 92-37;

C.~Kounnas, F.~Zwirner and I.~Pavel, Phys. Lett. {\bf B335}, 403 (1994);

P.~Binetruy and E.~Dudas, Phys. Lett. {\bf B338}, 23 (1994).
  
\bibitem{bims}A.~Brignole, L.~E.~Ib\'a\~nez, C.~Mu\~noz and
C.~Scheich, preprint FTUAM95/26.

\bibitem{ksyy}T.~Kobayashi, D.~Suematsu, K.~Yamada and Y.~Yamagishi,
Phys. Lett. {\bf B348}, 402 (1995).

\bibitem{dgt}S.~Dimopoulos, G.~F.~Giudice and N.~Tetradis, Nucl. Phys.
{\bf B454}, 59 (1995).

\bibitem{mv}S.~P.~Martin and M.~T.~Vaughn, Phys. Rev. {\bf D50}, 2282 (1994);

Y.~Yamada, Phys. Rev. {\bf D50}, 3373 (1994);

I.~Jack, D.~Jones, S.~Martin, V.~Vaughn and Y.~Yamada,
Phys. Rev. {\bf D50}, 5481 (1994).

\end{thebibliography}
\end{document}